\documentclass[aps,pre,twocolumn,superscriptaddress]{revtex4-2}
 \usepackage{subfigure}
 \usepackage{amssymb}
 \usepackage{amsfonts}
 \usepackage{amsmath}
 \usepackage{amsthm}
 \usepackage{epsfig}
 \usepackage{float}
 \usepackage[usenames,dvipsnames]{color}
 \usepackage{graphics,graphicx,xcolor,subfigure}

 \usepackage[hidelinks,unicode=true]{hyperref}
 \hypersetup{colorlinks=true,
 	linkcolor=blue,
 	urlcolor=blue,
 	citecolor=blue,
 	pdfhighlight=/N%
 } 
 
 \usepackage{comment}
 \usepackage[normalem]{ulem}
 \usepackage{microtype}

 \newcommand{\av}[1]{\langle {#1} \rangle}

 \newcommand{\kmax}{k_\text{max}}
 \newcommand{\kmin}{k_\text{min}}

 \newcommand{\Sp}{\text{S}_\text{p}}
 \newcommand{\Su}{\text{S}_\text{u}}
 \newcommand{\lbp}{\lambda_\text{p}}
 \newcommand{\lbu}{\lambda_\text{u}}
 \newcommand{\lbc}{\lambda_\text{c}}
 \newcommand{\Gammapu}[1]{\Gamma_{#1}^{\text{p}\rightarrow\text{u}}}
 \newcommand{\Gammaup}[1]{\Gamma_{#1}^{\text{u}\rightarrow\text{p}}}
 \newcommand{\putuj}[2]{{#1}^\text{(u)}_{#2}}
 \newcommand{\putpj}[2]{{#1}^\text{(p)}_{#2}}
 \newcommand{\dup}{d_\text{up}}
 \newcommand{\dpu}{d_\text{pu}}

\begin{document}
\title{Epidemic outbreaks with adaptive prevention on complex networks}

\author{Diogo H. Silva}\thanks{\textit{Present address}: Instituto de Ci\^{e}ncias Matem\'{a}ticas e de Computa\c{c}\~{a}o, Universidade de S\~{a}o Paulo, S\~{a}o Carlos, SP, Brazil.}
\affiliation{Departamento de F\'isica, PUC-RIO, Rua Marquês de São Vicente, 225, 22451-900, Rio de Janeiro, Brazil.}
\affiliation{National Institute of Science and Technology for Complex Systems, 22290-180, Rio de Janeiro, Brazil}

\author{Celia Anteneodo~}
\affiliation{Departamento de F\'isica, PUC-RIO, Rua Marquês de São Vicente, 225, 22451-900, Rio de Janeiro, Brazil.}
\affiliation{National Institute of Science and Technology for Complex Systems, 22290-180, Rio de Janeiro, Brazil}

\author{Silvio C. Ferreira}
\affiliation{Departamento de F\'{\i}sica, Universidade Federal de Vi\c{c}osa, 36570-900 Vi\c{c}osa, Minas Gerais, Brazil}
\affiliation{National Institute of Science and Technology for Complex Systems, 22290-180, Rio de Janeiro, Brazil}

\begin{abstract} 
The adoption of prophylaxis attitudes, such as social isolation and use of face masks, to mitigate epidemic outbreaks strongly depends on the support of the population. In this work, we investigate a susceptible-infected-recovered (SIR) epidemic model, in which the epidemiological perception of the environment can adapt the behavior of susceptible individuals towards preventive behavior. {Two compartments of susceptible individuals are considered, to distinguish those that adopt or not profilaxis attitudes.} Two rules, depending on local and global epidemic prevalence, for the spread of the epidemic in heterogeneous networks are investigated. We present the results of both heterogeneous mean-field theory and stochastic simulations. The former performs well for the global rule, but misses relevant outcomes of simulations in the local case.   In simulations,  only local awareness  can significantly raise the epidemic threshold,   delay the peak of prevalence,  and reduce the outbreak size. Interestingly, we observed that increasing the local perception rate leads to less individuals recruited to the protected state, but still enhances the effectiveness in mitigating the outbreak. We also report that network heterogeneity substantially reduces the efficacy of local awareness mechanisms since hubs, the super-spreaders of the SIR dynamics, are little responsive to epidemic environments in the low epidemic prevalence regime. Our results indicate that strategies that improve the perception of who is socially very active can improve the  mitigation of epidemic outbreaks.
\end{abstract}

\maketitle

\section{Introduction}

During the last two years facing a lethal pandemic, adoption of non-pharmaceutical interventions such as enhanced hygiene, face masks~\cite{Kwon2021},  physical distancing~\cite{Flaxman2020}, mobility restrictions~\cite{Chinazzi2020} were fundamental to mitigate the  virus transmission, while efficient vaccines and treatments were not available to everyone. Tools exploiting technological advances, such as contact-tracing apps~\cite{Ferretti2020,Cencetti2021} and GPS~\cite{Lucchini2021},  are allies to identify places of high epidemic incidence aiming at optimizing the non-pharmaceutical strategies. However, their efficiencies depend on population adhesion which is influenced by many factors such as the risk perception~\cite{Ferrer2018,Ferrer2015},  optimistic bias~\cite{Jocelyn2020}, government enforcement~\cite{Schmelze2021}, political lean~\cite{Schmelze2021,Frey2020}, social condition~\cite{vonWyl2021}, and so on.   

Understanding how individuals of the population respond to epidemiological environments during an epidemic outbreak is a challenge connecting epidemiology, sociology and medicine~\cite{Ferguson2007}. The connection among these areas can be seen, for example, in the spread of antivax content  in social media~\cite{Burki2019,Germani2021}, which plays an important role in the adhesion of consolidated vaccination campaigns such as measles and seasonal flu. In general, information changes epidemic spreading~\cite{HATZOPOULOS2011,KISS2010,Wu2012,Zhang2014,Granell2013,Granell2014,Paulo2019,LI2020}  while the effects depend on the nature of this information, for instance with regard to whether it is local (from near neighbors) or global (from the whole population). In both cases, the size of the outbreak is reduced~\cite{HATZOPOULOS2011,KISS2010,Wu2012,LI2020}, but only local awareness (or perception) is capable of raising the epidemic threshold~\cite{HATZOPOULOS2011,Wu2012,Zhang2014}. The spreading of a large amount of content about the ongoing COVID-19 pandemics in social media has raised questions about whether or not phenomena such as echo chambers, in which communication is highly limited to a group of people who share similar views on a topic, might influence the prophylaxis behavior of these individuals~\cite{Burki2019, valensise2021, Pires2021}.
  
Adaptive human behaviors in response to available information during an epidemic outbreak can be modeled exploiting different approaches. A recurrent one is using  multiplex networks~\cite{Granell2013,Paulo2019,Granell2014,Wang2016}, in which information and epidemic spread in interconnected networks which represent different social and physical contacts. See Ref~\cite{Bedson2021} for other approaches  to model the behavioral adaptation associated with epidemics. The capacity of awareness in altering the shape of epidemic outbreaks is well grounded in different theoretical approaches~\cite{Bedson2021}.  However, there  are points not fully understood and the topic remains very active~\cite{LI2020,Marco2020,Steinegger2020,Pires2021,Zhao2021}. We contribute to this field addressing, from a theoretical point of view, the important question of how to efficiently mitigate an epidemic outbreak under reduced socioeconomic costs.  We investigate an epidemic spreading with adaptive behavior in response to either local or global perception of epidemiological indicators. 
The model consists of a susceptible-infected-recovered (SIR)~\cite{Moreno2002}  epidemic process including two compartments of susceptible individuals, who adopt or not prophylaxis to reduce the probability of contagion. The former group is infected with a rate smaller than the latter. In addition, the individuals can switch their behavior from protected to unprotected, depending on  either local or global perception of the epidemic prevalence. 
We tackle the problem both analytically and by extensive stochastic simulations using a single layer static network in order to make the problem as simple as possible, but preserving the contact heterogeneity, central for epidemic spreading~\cite{Satorras2002,Newman2002}. Power-law networks are analyzed, as well as random regular networks for comparison. We confirm that both  local and global awareness can reduce the size of the epidemic outbreak while only the former is able to raise the epidemic threshold as previously observed elsewhere~\cite{HATZOPOULOS2011, Wu2012, Zhang2014}. However, we also report a complex interplay between local awareness and response to epidemiological environment and the number of individuals moved into protected compartment: we observed a wide range of parameters for which a higher rate of moving to the protected state implies that fewer individuals have to do that for an efficient mitigation of the outbreak. This behavior, observed in simulations, cannot be reproduced under the population mixing hypothesis used in the  mean-field approaches. The underlying mechanisms behind these findings are rationalized. Another important result we report is that heterogeneity of the contact network reduces the efficacy of the local awareness mechanism since the hubs, which are the super-spreaders of the SIR dynamics, have low perception of the epidemic prevalence due to high number of contacts.

The rest of this paper is organized as follows. In Sec.~\ref{sec:model}, we describe the model of  adaptive behavior in response to epidemiological perception.  The heterogeneous mean-field theory and their results are presented in Sec.~\ref{sec:theories}. Stochastic simulations are presented,  discussed and compared with theoretical predictions in Sec.~\ref{sec:result}. Finally, our main results and conclusions are summarized in Sec.~\ref{sec:conclusions}. One appendix with the computer implementation of the stochastic simulations ends the paper.

\section{Model} 
\label{sec:model}

We consider an epidemic process where the individuals of a population are represented by nodes of an undirected graph, with $i=1,\ldots,N$ nodes and connections encoded by the adjacency matrix $A_{ij}$, which assumes binary values $A_{ij}=1$ if the node $i$ and $j$  are connected and $A_{ij}=0$, otherwise. We investigate the role of protection due to local and global environment perceptions, using a modified SIR dynamics with two compartments of susceptible  individuals, $\Sp$ and $\Su$, which may adopt or not self-protection, respectively. The infection rate per contact of an unprotected susceptible individual is $\lbu \equiv\lambda$  while for a protected one it is $\lbp \equiv\alpha\lambda$,  where $0\leq\alpha< 1$ represents the increased protection of the $\Sp$ compartment with respect to the $\Su$ compartment. Susceptible individuals can change their states from protected to unprotected, and vice-versa, influenced by the perception of either local or global epidemic prevalence. Initially, the whole population of susceptible individuals belongs to $\Su$. 

Empirical data gathered during the COVID-19 pandemics~\cite{Muscillo2020} suggest that  more socially active persons are more reluctant to reduce their social interactions and tend to leave quarantines earlier~\cite{DeMeijere2021}. To incorporate this feature in the model with local information, we consider that a $\Su$ individual $i$ changes its behavior and moves to the $\Sp$ compartment with rate 
\begin{equation}
\Gammaup{i}=\dup\sum_{j}\frac{A_{ij}}{k_{i}}\sigma_{j},
\label{eq:dif_local_a}
\end{equation}
in which,  $k_{i}$ is the degree of node $i$, $\sigma_{i}=1$ if node $i$ is infected and $\sigma_i=0$ otherwise. Conversely, a protected susceptible individual becomes unprotected ($\Sp\rightarrow\Su$) with rate 
\begin{equation}
\Gammapu{i}=\dpu\sum_{j}\frac{A_{ij}}{k_{i}}(1-\sigma_{j}).
\label{eq:dif_local_b}
\end{equation}
Note that the transition rates given by Eqs.~\eqref{eq:dif_local_a} and~\eqref{eq:dif_local_b} are determined by local information  and each infected neighbor is less relevant for more connected individuals. The parameters $\dup$ and $\dpu$ are the perception rates at which behavior switches from unprotected to protected and vice-versa, respectively, in response to the local environment.

The global awareness variant of the model considers the total fraction of infected individuals. The transition rates for $\Su\rightarrow\Sp$ and $\Sp\rightarrow\Su$ become
\begin{equation}
\Gammaup{i}=\dup\rho, 
\label{eq:dif_global_a}
\end{equation}
and
\begin{equation}
\Gammapu{i}=\dpu(1-\rho),
\label{eq:dif_global_b}
\end{equation}
respectively. Similarly, $\dup$ and $\dpu$ are now the perception rates of the global epidemic prevalence.  Despite the fact that velocity of information {propagation} influences the epidemic prevalence~\cite{Paulo2019}, we  assume that the infectious state is instantaneously perceptible under both  global and local rules. This is a realistic assumption, for instance when daily updates of the number of cases and deaths are published, as currently occurs in the COVID19 pandemic.

The model  presents upper and lower bounds for the epidemic threshold given by  $\lambda^\text{SIR}_\text{c}\leq\lbc\leq\alpha^{-1}\lambda^\text{SIR}_\text{c}$. The lower bound  corresponds to $\dup\rightarrow0$, in which case unprotected individuals do not change their behavior, while the upper bound corresponds to $\dup\rightarrow\infty$,  when individuals instantaneously adopt  self-protection behavior if infected people are perceived, either in the neighborhood for the local rule or in the whole system for the global one. These bounds are not affected by $\dpu$.

\section{Heterogeneous mean-field theory}
\label{sec:theories}

In the heterogeneous mean-field (HMF) theory~\cite{Romualdo2001_a, Romualdo2001_b,Romualdo2002}, the set of nodes of the same degree are assumed to be statistically equivalent. The adjacency matrix is replaced by the probability  that a node of degree $k$ is connected to nodes of degree $k'$, which is given by the connectivity matrix $C_{kk'}=kP(k'|k)$~\cite{Boguna2003}, where  $P(k'|k)$ is the conditional probability that a neighbor of a node of degree $k$ has degree $k'$. The infectious status is replaced by the probability $\rho_k$ that a node of degree $k$ is infected. The protection rates defined in Eqs.~\eqref{eq:dif_local_a} and \eqref{eq:dif_local_b}, now for a node of degree $k$, become \begin{equation}
\Gammaup{k}=
\frac{\dup}{k}\sum_{k'} kP(k'|k)\rho_{k'} = \dup\Phi_k 
\end{equation}
and
\begin{equation}
\Gammapu{k}= \dpu\left(1-\Phi_k\right).
\label{eq:HMF_L2}
\end{equation}
In the special and useful case of uncorrelated networks, we have $P(k'|k)=k'P(k')/\av{k}$~\cite{Boguna2004} such that $\Phi_k=\Phi$ is independent of the degree. For the global awareness model, Eqs.~\eqref{eq:dif_global_a} and \eqref{eq:dif_global_b} are directly evaluated since $\rho=\sum_k\rho_k P(k)$.

Following standard HMF theory for the SIR dynamics on networks~\cite{Moreno2002,Boguna2003}, the temporal evolution of the prevalence in the infected ($\rho_k$), recovered ($r_k$), and susceptible ($\putuj{s}{k}$ and  $\putpj{s}{k}$) compartments is given by
\begin{eqnarray}
\frac{d\rho_{k}}{dt}&=&-\mu\rho_{k}+k \left(\lbu \putuj{s}{k}+\lbp \putpj{s}{k} \right)\Theta_{k}\label{eq:HMF1_a},\\
\frac{d \putuj{s}{k}}{dt}&=&-  \lbu k \putuj{s}{k} \Theta_{k}+\Gammapu{k} \putpj{s}{k} - \Gammaup{k} \putuj{s}{k} \label{eq:HMF1_b}, \\
\frac{d \putpj{s}{k}}{dt}&=&-  \lbp k \putpj{s}{k} \Theta_{k}+\Gammaup{k} \putuj{s}{k} - \Gammapu{k} \putpj{s}{k} \label{eq:HMF1_c},
\end{eqnarray}
which form a  closed system with the normalization condition $\rho_{k}+r_{k}+\putpj{s}{k}+\putuj{s}{k}=1$, {and $\mu$ is the recovery rate.} The quantity $\Theta_{k}$ is the probability that a link selected at random points to an infected node, and it is given by 
\begin{eqnarray}
\Theta_{k} = \Theta=\sum_{k'}\frac{(k'-1)P(k')}{\langle k\rangle}\rho_{k'},
\end{eqnarray} 
in the absence of degree  correlations~\cite{Boguna2003}. 

\begin{figure}[h]		
	\includegraphics[width=0.98\linewidth]{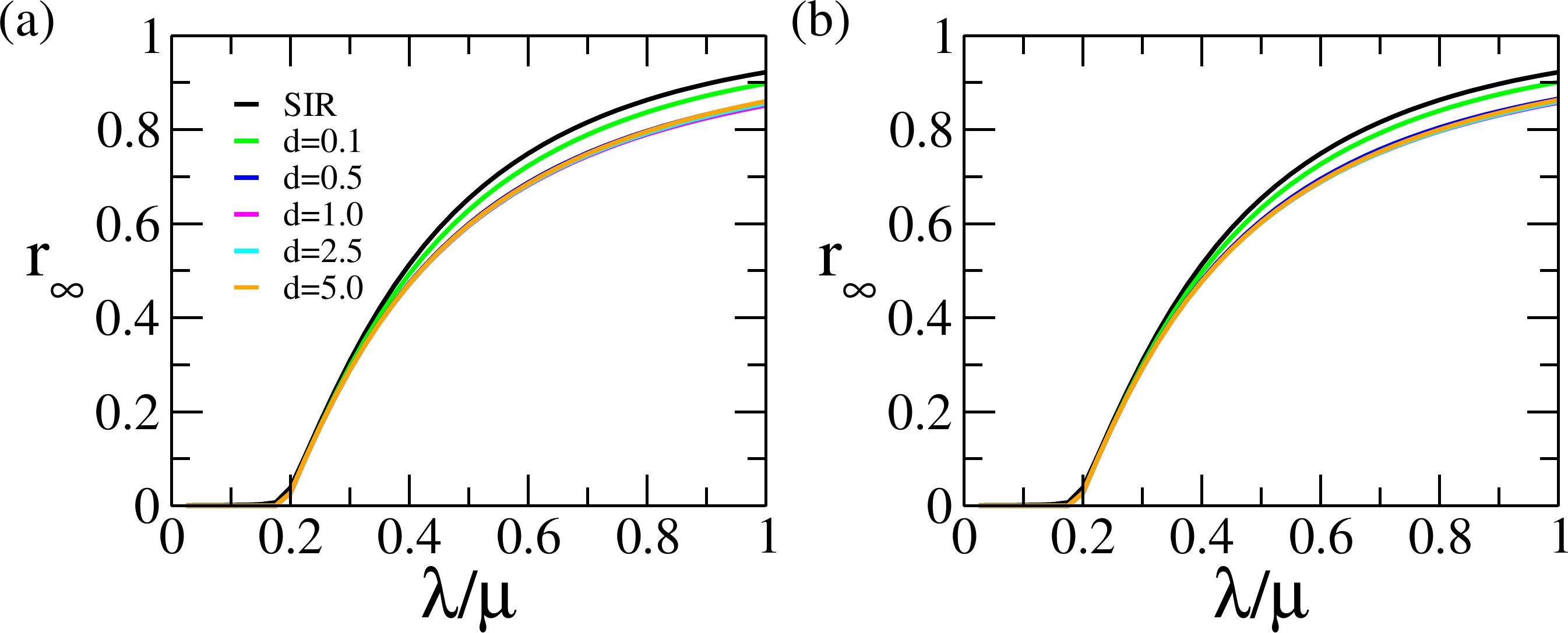}\\
	\includegraphics[width=1\linewidth]{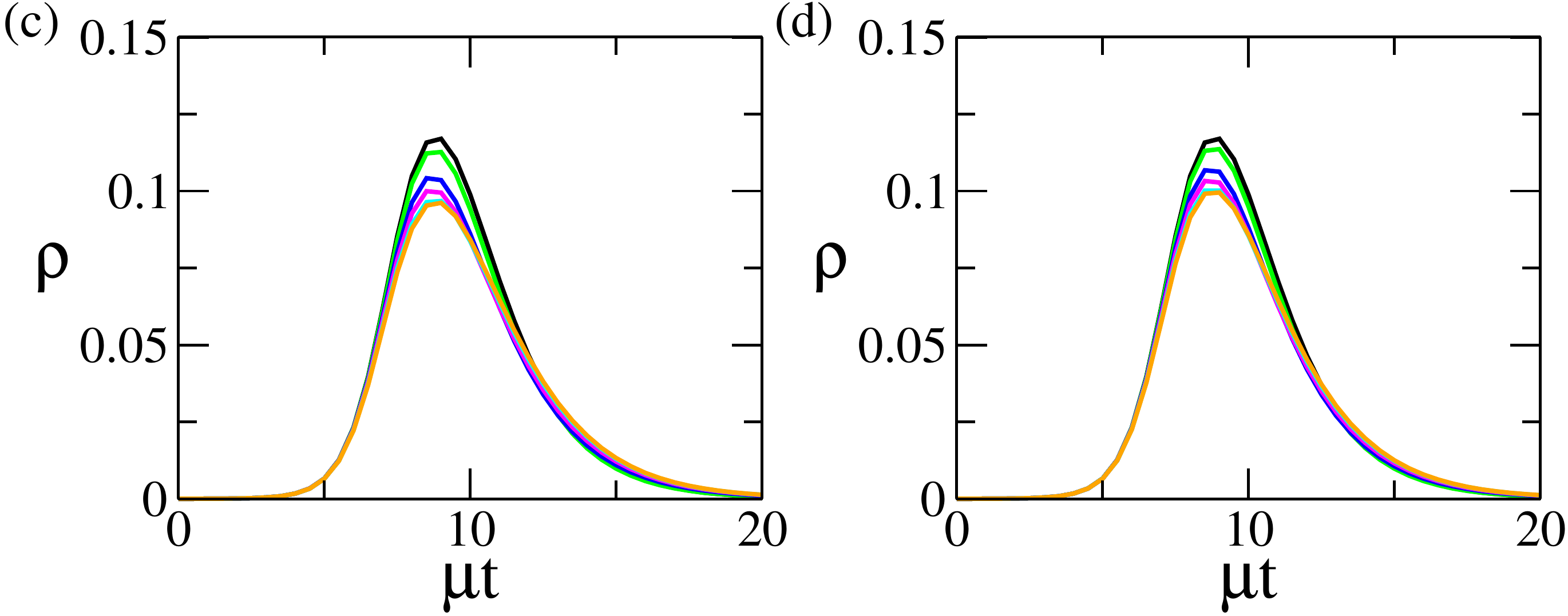}\\
	\includegraphics[width=1\linewidth]{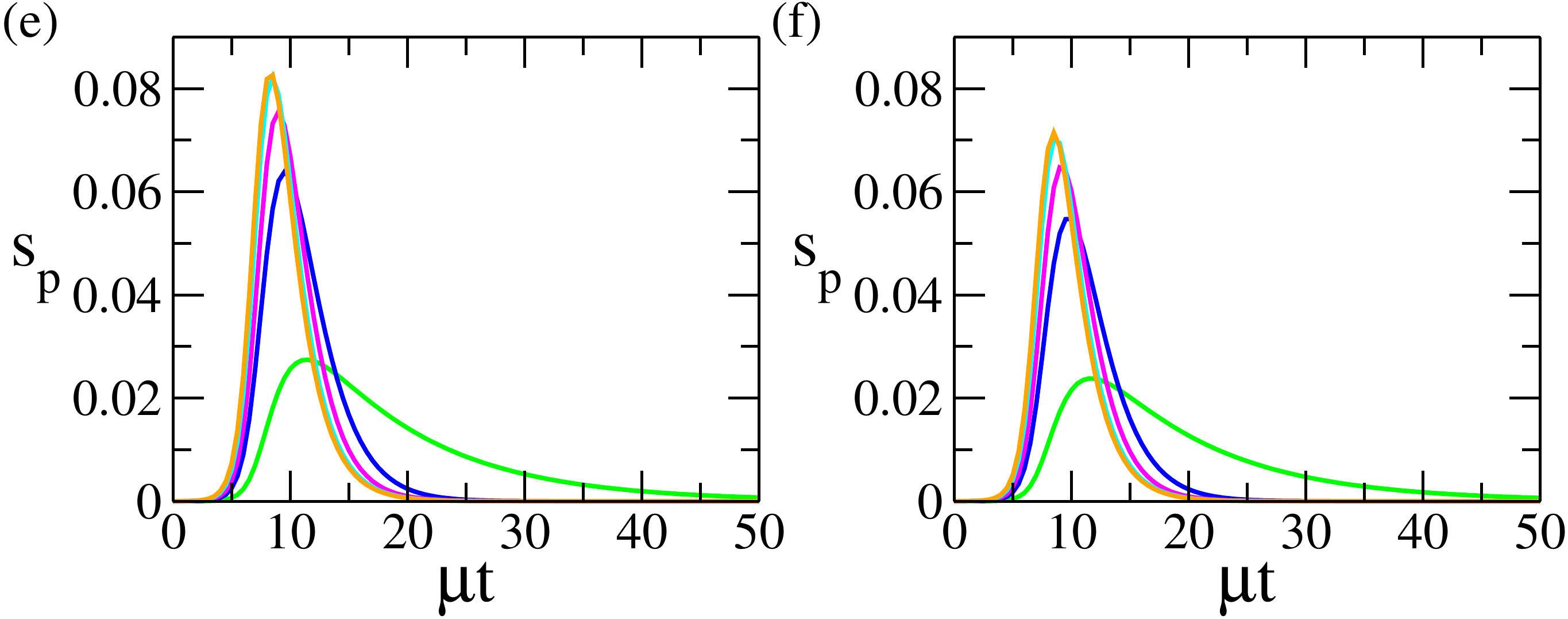}		
	\caption{HMF results for local (a,c,e)  and global (b,d,f) perception rules, from the numerical integration of Eqs.~(\ref{eq:HMF1_a})-(\ref{eq:HMF1_c}). (a,b) Outbreak size as a function of $\lambda_\text{u}=\lambda$ with $\lambda_\text{p}=0$, (c,d) epidemic prevalence, (e,f) protected population as functions of time for fixed $\lambda=0.46\mu>\lbc$.  The values of  $d=\dup=\dpu$,  which control the protection rates, are given in the legend (the standard SIR is recovered when $d\equiv 0$). Power-law networks with degree distribution $P(k)\propto k^{-\gamma}$, $ k_{0}\le k\le k_\text{c}$, $\gamma=3.5$, $N=10^{5}$, $k_{0}=3$ and $k_\text{c}\sim N^{1/\gamma}$ were used. 
	\label{fig:HMF_LOC_GLO}
	}
\end{figure}

A stability analysis around the fixed point $(\rho_{k},r_{k},\putpj{s}{k},\putuj{s}{k})=(0,0,0,1)$ leads to the same epidemic threshold of the standard SIR model on networks~\cite{Moreno2002}, given by
\begin{equation} \label{eq:lbc}
\lbc = \frac{\av{k} \mu}{\av{k^2}-\av{k}},
\end{equation}
implying that protection due to perception does not play a role in the onset of the epidemic outbreak in neither local nor global rules. This is in contrast with the reduction of the epidemic threshold reported for modified SIR models with other {nonlinear degree} dependencies of the rates {on awareness}~\cite{Wu2012,Zhang2014}. The independence predicted by Eq.~(\ref{eq:lbc}) is verified in the numerical integration of the HMF equations, for local and global rules, {as shown in Fig.~\ref{fig:HMF_LOC_GLO} for several values of  $d\equiv\dup=\dpu$}, and $\alpha=0$ ({meaning that} protected individuals cannot be infected) for a power-law degree distribution (PL), $P(k)\sim k^{-\gamma}$ with $\gamma=3.5$, $N=10^{5}$, and lower degree $k_\text{0}=3$ and upper cutoff given by and   $k_{\text{c}}\sim N^{1/\gamma}$.

{Moreover, for rates $\dup=\dpu$, the final fraction of recovered individuals, shown in Figs.~\ref{fig:HMF_LOC_GLO}(a) and (b), is  not significantly  influenced by the introduction of  local or global rules, varying little with $d$.  The prevalence peak slightly decreases with $d$  (Figs.~\ref{fig:HMF_LOC_GLO}(c)-(d).) Since the rates $\dup$ and $\dpu$ have similar values and increase concomitantly, the larger number of individuals recruited to the $\Sp$ compartment 
is compensated by the shorter time they stay protected and the net result is a little dependence of the epidemic curves on $\dup=d$. Meanwhile, the peak of the protected population increases with $d$, as expected (Figs.~\ref{fig:HMF_LOC_GLO}(e)-(f)). 
We observe that the impact of local and global rules is very similar, weak and only slightly more beneficial when rules are local.} 
This picture is qualitatively representative of the more general case $\dpu  \ge \dup>0$, which means protected individuals becoming unprotected  faster than the converse transition. Both epidemic curves approach the SIR limit if $\dpu\gg \dup$.

\begin{figure}[h!]
	\includegraphics[width=1\linewidth]{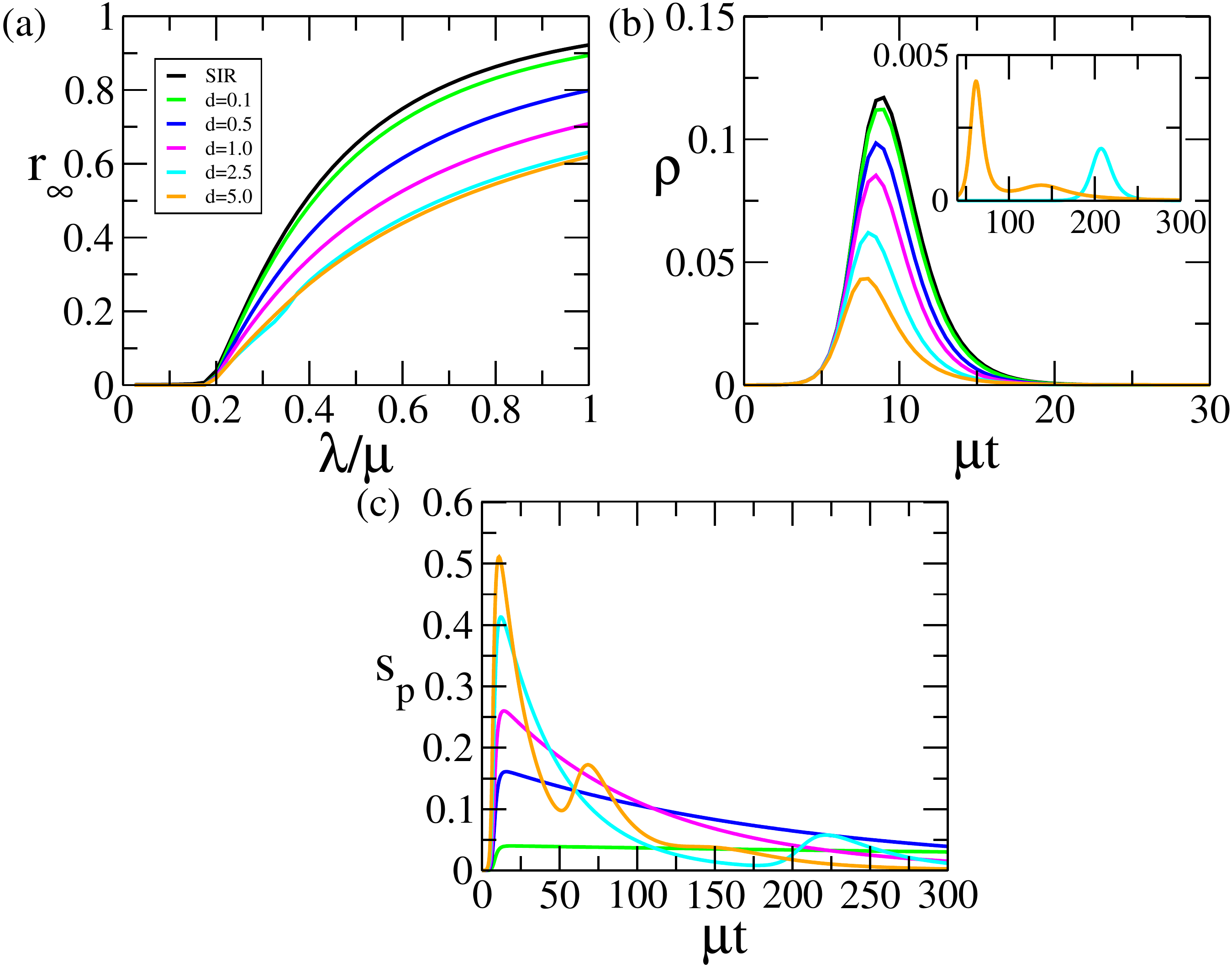}\\	
	\caption{HMF results for local perception rules, in the extreme case $\lambda_\text{p}=0$, and different values of  $d=\dup=100 \dpu$ given in the legend. (a) Outbreak size as a function of $\lambda=\lambda_\text{u}$.  (b) Epidemic prevalence and (c) fraction of protected individuals as function of time,  for $\lambda=0.46\mu$. The inset in (b) shows a zoom of the epidemic prevalence for longer times.  The integration was performed in PL networks with $\gamma=3.5$, $k_\text{c}\sim N^{1/\gamma}$ and $N=10^{5}$. }
	\label{fig:HMF_LOCAL2a}
\end{figure}

Differently, if $\dup>\dpu$, the outcomes depend significantly on the perception rates, as can be seen in  Fig.~\ref{fig:HMF_LOCAL2a}, for $\dup=100 \dpu$, chosen to make the effect more evident. Also in this regime, only subtle  differences are observed between local and global perception and then only the local case is shown. The outbreak size for different values of $d$ is shown in  Fig.~\ref{fig:HMF_LOCAL2a}(a). In general, for  $\dup\gg \dpu$,  once an unprotected individual moves to the $\Sp$ compartment, she or he can stay at this condition for periods longer than the outbreak duration and never be infected. These protected individuals are comparable to vaccinated ones with waning immunity, which take the vaccine decision based on its perception of the epidemic scenario and this immunity wanes after a characteristic time smaller than the total time for epidemic eradication.  

Still in the regime $\dup\gg\dpu$ with a high perception rate ($\dup/\mu\ge 2.5$ in Fig.~\ref{fig:HMF_LOCAL2a}), both the epidemic prevalence and fraction of protected individuals exhibit damped oscillations forming small subsequent outbreaks since some protected individuals change their behavior before epidemic eradication, as shown in Fig.~\ref{fig:HMF_LOCAL2a}(c) and the inset of Fig.~\ref{fig:HMF_LOCAL2a}(b). For the global rule (not shown), the resurgence is more damped and is delayed, with respect to the local case. The mechanism behind this phenomenon is analogous to a non-coexisting prey-predator dynamics. 

We also investigated the case $\dup=4\dpu$. While oscillations were not observed, the other conclusions remain valid. Actually, oscillations were not observed in any stochastic simulation we have performed (see Sec.~\ref{sec:result}). Let us note that the resurgence of the epidemic outbreak in the case $\dup=100\dpu$ started from a tiny fraction of infected individuals observable only in the deterministic equations since fluctuations would push the system toward the absorbing state in such a regime.

\section{Numerical simulations}
\label{sec:result}

We ran stochastic simulations on networks following the algorithm detailed in Appendix~\ref{app:algo}. We consider the limit case  $\alpha=0$, in which susceptible individuals in $\Sp$ compartment cannot be infected.

\subsection{Random regular networks} 

\begin{figure}[h!]
\includegraphics[width=0.98\linewidth]{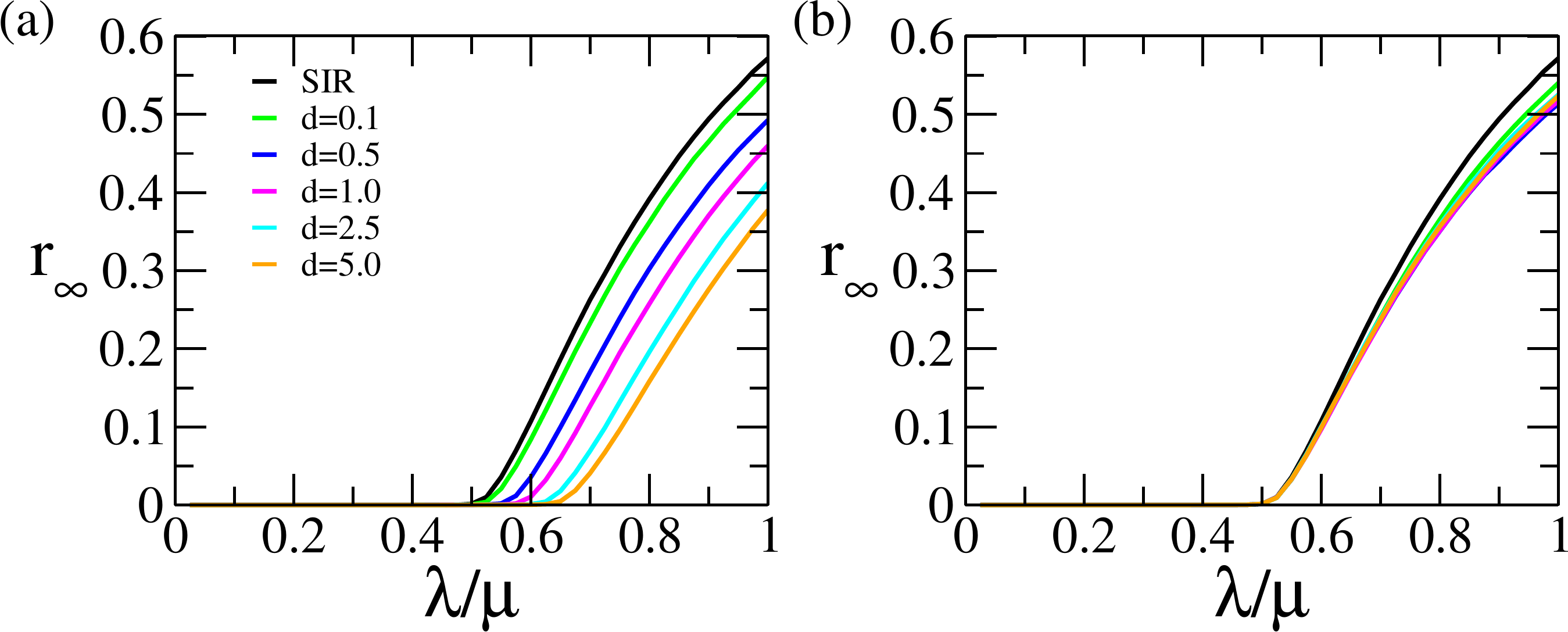}\\
\includegraphics[width=1\linewidth]{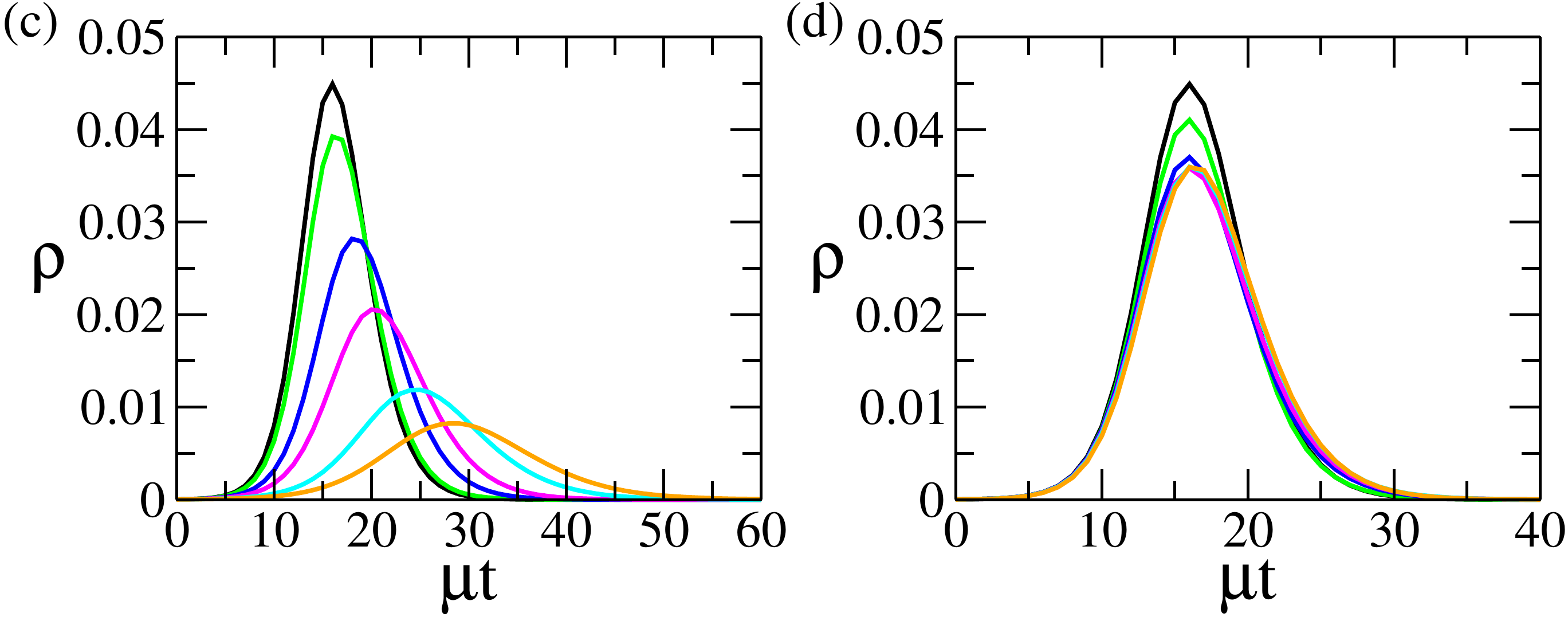}	 
\includegraphics[width=1\linewidth]{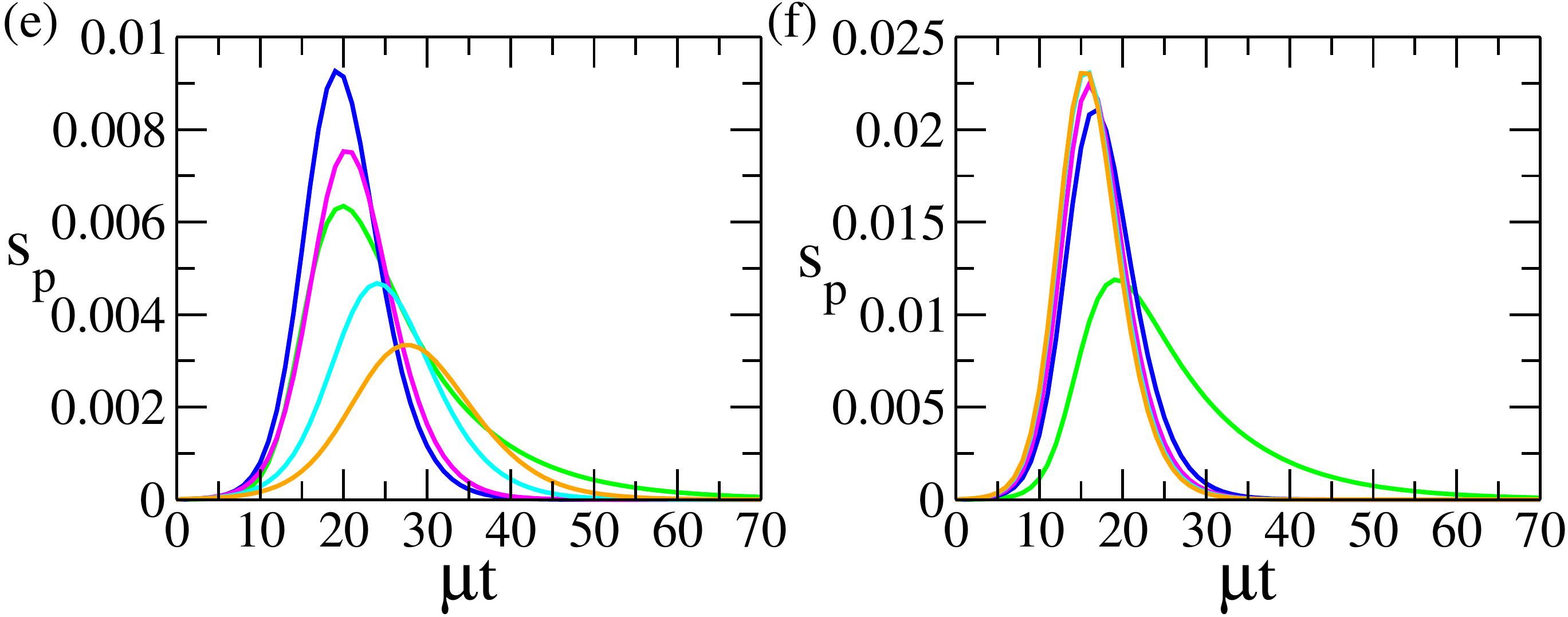}
	\caption{ Stochastic simulations of the  SIR dynamics with  (a,c,e) local and   (b,d,f) global  awareness rules in homogeneous networks. (a,b) Outbreak size as a function of the infection rate, and  (c,d) epidemic prevalence and (e,f) density of protected individuals as a function of time. The reduction factor is $\alpha=0$ and the rate of transition between protected and unprotected susceptible subpopulations are $d=\dup=\dpu$ with the values of $d$ indicated in the legend. In (c,d,e,f), we adopted a fixed value $\lambda_\text{u}=\lambda=0.8>\lbc$. The results were obtained in  RR networks  with $k=4$  and $N=10^{5}$ nodes.} 
	\label{fig:SIR_RRN}
\end{figure}

We start our analysis with the homogeneous case of random regular (RR) networks, where all nodes have the same degree and connections are performed at random~\cite{Ferreira2012}. Both local and global awareness reduce the size of the epidemic outbreak if compared with standard SIR, while only the local rule raises significantly the epidemic threshold, as shown in Figs.~\ref{fig:SIR_RRN}(a) and (b). 

For the fixed value of $\lambda_\text{u}=\lambda=0.8\mu>\lbc$, Figs.~\ref{fig:SIR_RRN} (c) and (d) show that the peak of prevalence is not significantly influenced by raising $\dup=\dpu$ for global awareness, but it is for the local case. Local perception promotes the flattening of the epidemic curve with delayed and less pronounced peaks of prevalence, while the delay is negligible in the global rule. Flattening the curve is widely sought in emerging disease outbreaks to avoid overloading healthcare systems and to provide more time to develop treatments or vaccines.
Notice, however, that the peak of susceptible individuals in the protected compartment $\Sp$, shown in Figs.~\ref{fig:SIR_RRN} (e) and (f), is also reduced as the awareness rate increases in the local case. This result is counter-intuitive from the mean-field perspective since a higher responsiveness to epidemic prevalence is  expected to increase the number of individuals in the protected compartment when perception rates to enter and to leave $\Sp$ are symmetric,  as observed in the  HMF results shown in Fig.~\ref{fig:HMF_LOC_GLO}. However, Figure~\ref{fig:SIR_RRN}(e) indicates a different interpretation where the  quicker a local response is, the  transmission is more efficiently blocked and, consequently, less individuals are sent to the protected compartment. Thinking about quarantine protocols, our results point out that the more rigorous the criteria for adopting quarantine, the smaller the number of individuals who need to be isolated and leave their work or study spots. This is socioeconomically {worrysome} since disadvantaged socioeconomic groups {might not be able to remain}  isolated for long periods~\cite{Chang2021}  and determining optimal time to implement restrictions  is a fundamental issue~\cite{Thompson2018,Morris2021,Di_Lauro2021}. 

The outcomes of simulations for symmetrical perception,  when rules are local, reveals relevant aspects not captured by the HMF theory, which assumes well-mixed populations and is widely adopted in epidemiological modeling~\cite{keeling2008modeling}. A first one is the dependence of the epidemic threshold on the perception rates. A second one is the reduction of the susceptible population which migrates to the protected compartment as the epidemic perception rate is increased, but still flattening the curves and mitigating the impacts of the outbreak.  A last one is that epidemic curves in simulations are flattened as the perception rate increases, differently to what is predicted by the theory.  However, for global awareness, the HMF results remain essentially valid.
\subsection{Power-law networks}

To investigate the role of heterogeneity, we consider synthetic uncorrelated networks with power-law degree distribution, $P(k)\sim k^{-\gamma}$, generated with the uncorrelated configuration model (UCM)~\cite{Boguna2004,Catanzaro2005}, in which  self and multiple connections are forbidden. We considered $\gamma=3.5$ and an upper cutoff for degree given by  $P(k_\text{c})N=1$, hence $k_{\text{c}} \sim N^{1/\gamma}$. The choice of  $\gamma>3$ leads to a finite epidemic threshold for the pure SIR dynamics in contrast with an asymptotically null threshold of SIR on scale-free networks with $\gamma<3$~\cite{Moreno2002}. We fixed the network size to $N=10^5$ implying an average degree $\av{k}=4.27$ very close to the $k=4$ considered in the homogeneous case.

\begin{figure}[h!]
	\includegraphics[width=\linewidth]{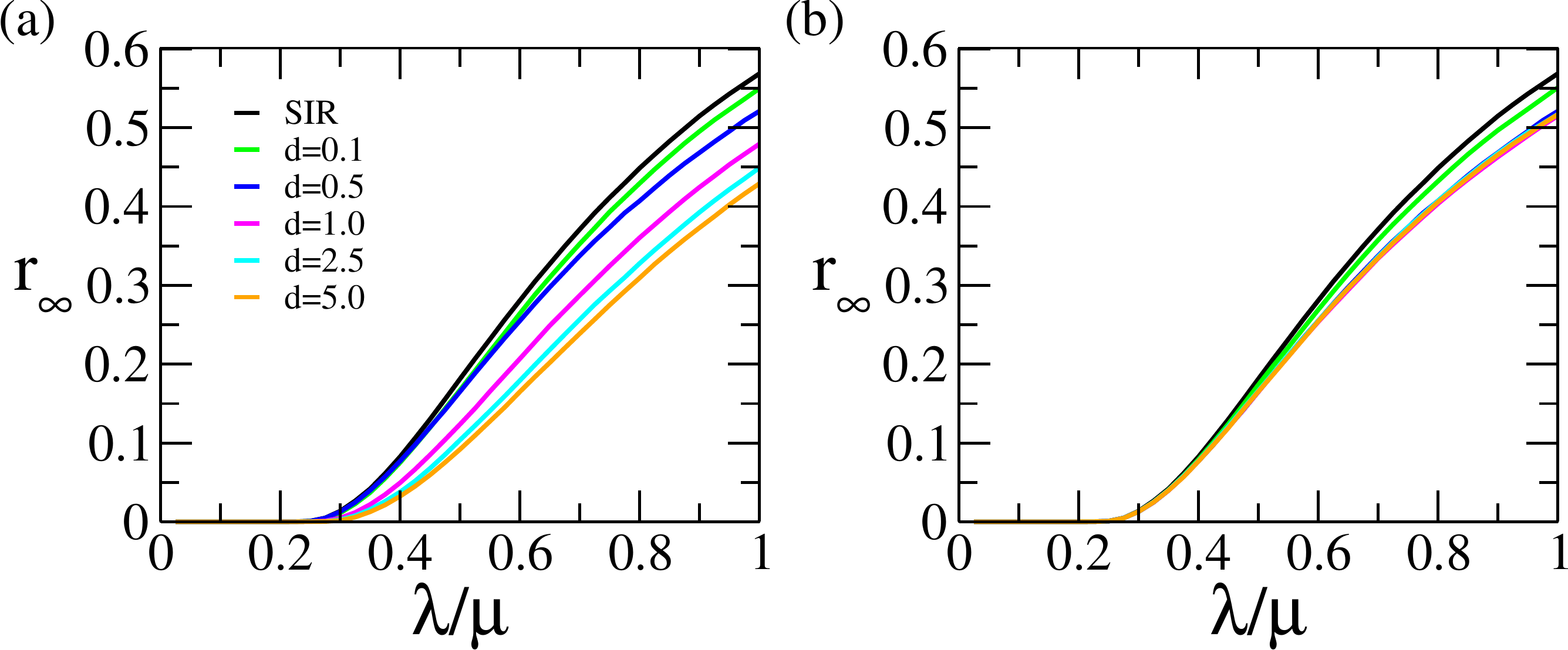}
	\caption{Outbreak size as a function of infection rate for $\alpha=0$ and different values of the transition rate between susceptible subpopulations $d=\dup=\dpu$ indicated in the legend for the modified SIR model with (a) local and (b) global awareness rules. Power-law networks  with $\gamma=3.5$ and $k_\text{c}\sim N^{1/\gamma}$, $N=10^{5}$, and average degree $\av{k}=4.27$ were considered.}
	\label{fig:SIR_PL}
\end{figure}

Heterogeneity implies the presence of super-spreaders, which in the case of the SIR dynamics are nodes of degree $k\gg \av{k}$ that promote the epidemic spreading and reduce the epidemic threshold for $\gamma=3.5$ when compared with the case of RR networks of similar average degree. The influence  of local and global rules in the epidemic threshold is similar to the homogeneous case, as shown in Fig.~\ref{fig:SIR_PL}, where we note that the threshold depends on the awareness rate in the local rule but not in the global one. Since heterogeneity of contacts significantly impacts the local epidemic perception, we hereafter discuss only the local information rule, since the global one is very similar to the homogeneous case with the same average degree.

\begin{figure}[th]
	\centering
	\includegraphics[width=0.7\linewidth]{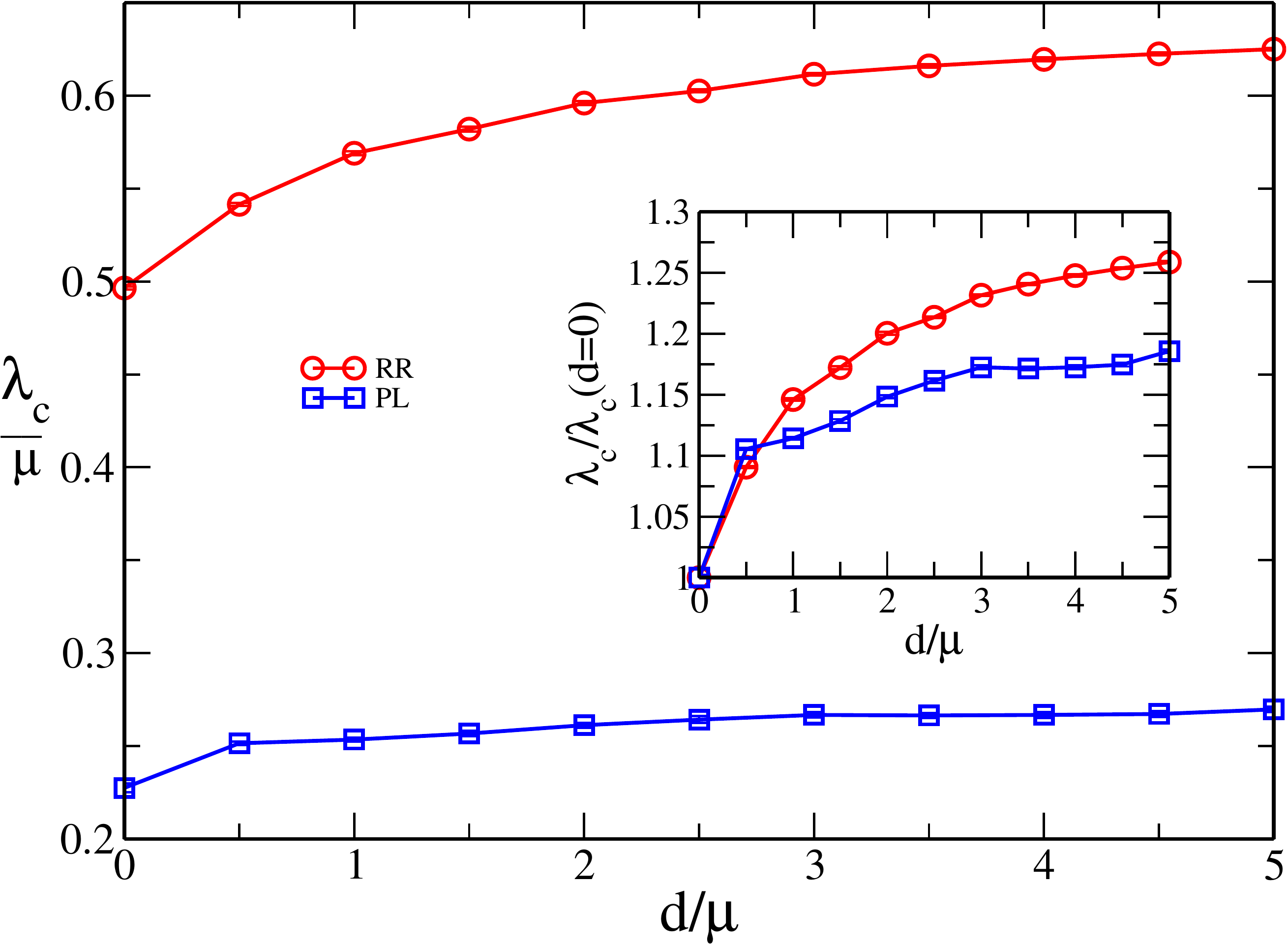}
	\caption{Epidemic threshold against awareness rate for local rule and epidemic spreading on RR ($\av{k}=4$) and PL ($\av{k}=4.27$ and $\gamma=3.5$) networks. The parameters $\alpha=0$, $d=\dup=\dpu$ were used in both cases. The inset shows the epidemic threshold scaled by the corresponding value of the SIR model.}
	\label{fig:limiarxdavgsuscalp0plxrrn}
\end{figure}

The epidemic threshold can be estimated considering variability of the epidemic outbreak defined as~\cite{Shu2015}
\begin{equation}
\Delta=\frac{\sqrt{\langle R^{2}_\infty\rangle-\langle R_\infty\rangle^{2}}}{\av{R_\infty}},
\end{equation}
where $R_\infty$ is the final outbreak size computed over independent runs on a network configuration. Figure~\ref{fig:limiarxdavgsuscalp0plxrrn} shows how the epidemic threshold depends on the local awareness rate for both RR and PL networks.  In both cases the threshold increases with awareness. The heterogeneity not only reduces the epidemic threshold, but also makes the awareness less effective to contain epidemic spreading expressed as a less significant raising of the threshold as a function of $\dpu=\dup=d$.  Actually, the behavioral responses caused by the local awareness rule, Eqs.~\eqref{eq:dif_local_a} and~\eqref{eq:dif_local_b}, imply that  hubs adopt protection with low probability  at low prevalence regime (many contacts and only a few infected ones), while quitting this behavior very quickly, as recently observed in the COVID19 pandemics~\cite{Muscillo2020,Marco2020}. 
Therefore, the dynamics of super-spreaders would not be drastically modified, implying a lower benefit of the local awareness when compared with the homogeneous counterpart of similar average degree. The case $\dpu>\dup$ will be considered in the sequence of this paper.

Heterogeneity also changes the shape of epidemic curves during an outbreak. Two basic quantities are the value of the prevalence peak $\rho_{\text{max}}$ and the outbreak duration $\tau$. The latter can be estimated as the curve width computed as 
\begin{equation}
\tau^2 = \int_{0}^{\infty} t^2 \phi(t)dt\,-\left[\int_{0}^{\infty} t \phi(t) dt\,\right]^2,
\end{equation}
where $\phi$ is the normalized epidemic curve given by
\begin{equation}
\phi(t) =\frac{\rho(t)}{\int_{0}^{\infty}\rho(t)dt}.
\end{equation}
In order to compare epidemic outbreaks ($\lambda>\lbc$) of different network structures we choose  fixing the basic reproductive number $R_0$ defined as the number of secondary infections generated by a single infectious individual introduced in a totally susceptible population~\cite{keeling2008modeling}. For the HMF theory of SIR dynamics on complex networks, we have~\cite{Newman2005}
\begin{equation}
R_0^\text{(SIR)}=\frac{\lambda}{\mu} \frac{\av{k^2}-\av{k}}{\av{k}}.
\end{equation}
Since $\Sp$ compartment is empty at $t=0$,  and the acquired awareness does not change $R_0^\text{(SIR)}$~\cite{LI2020}, it can be adopted to fix $\lambda$ for the general case with $d>0$ given that $\lambda>\lbc$. Numerical simulations corroborate this expectation. 

\begin{figure}[h]
	\centering
    \includegraphics[width=1\linewidth]{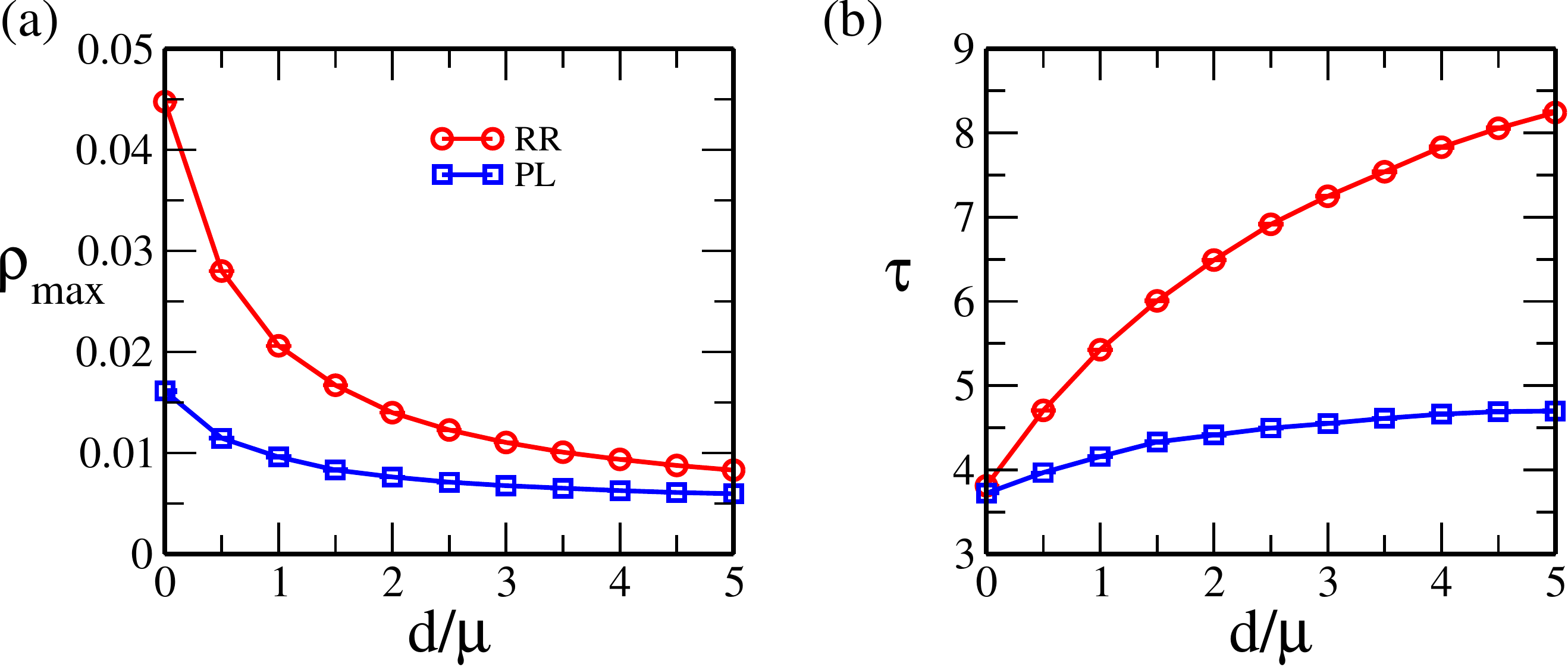}
	\caption{Comparison of (a) prevalence peak $\rho_{\text{max}}$ and (b) outbreak duration $\tau$  as function of the awareness rate in different networks for a fixed value of the basic reproductive number  $R_0^\text{(SIR)}=2.4$, which correspond to $\lambda=0.8$ for RR ($k=4$) and $\lambda=0.46$ for PL networks ($\gamma=3.5, \kmin=3$, $\kmax\sim N^{1/\gamma}$, $\av{k}=4.27$). The parameters $\alpha=0$ and $d=\dup=\dpu$ were used in both cases.}
	\label{fig:rhomaxtau}
\end{figure}

Figure~\ref{fig:rhomaxtau} shows that, while both the prevalence peak and the outbreak duration are lower in  heterogeneous networks in comparison with their homogeneous counterparts, the effect of increasing the awareness rate is much more pronounced in the homogeneous case due to the same reason that the epidemic threshold is also more affected, as discussed in Fig.~\ref{fig:limiarxdavgsuscalp0plxrrn}. Perception leads to more than five-fold ($\approx 5.4 \times$)  reduction in the epidemic peak and a twofold increase of the outbreak duration for the homogeneous case with $d=5$ against  more modest reduction of approximately three-fold ($\approx 2.7 \times$) in $\rho_{\text{max}}$  and increase of 26\% in $\tau$ for PL networks. 

\begin{figure}[th]
\includegraphics[width=\linewidth]{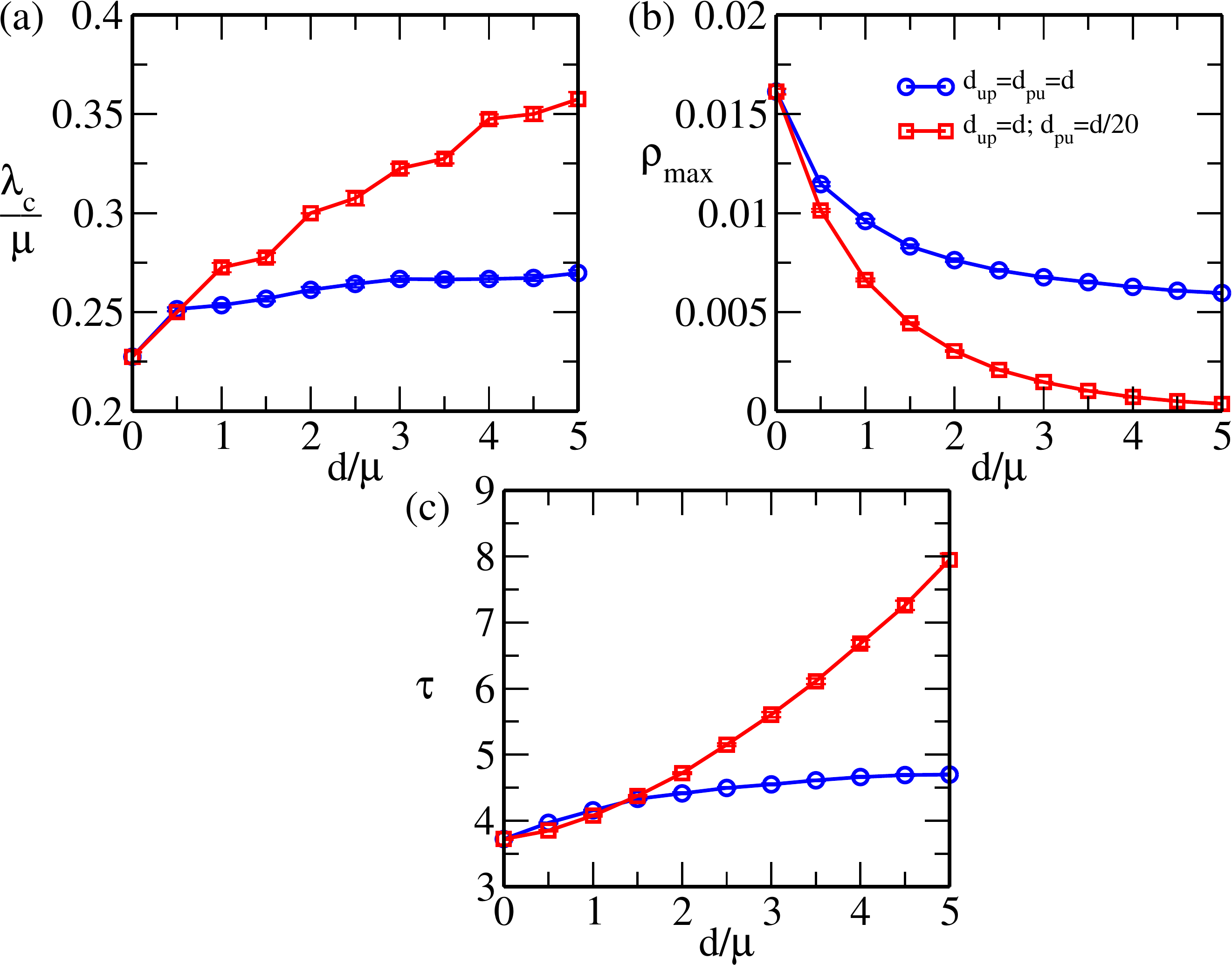}
	\caption{(a) Epidemic threshold, (b) peak prevalence and (c) outbreak duration as a function of the awareness rate. In (b) and (c),  $\lambda$ is chosen to provide the same basic reproductive number $R_0^\text{(SIR)}=2.4$. We consider the cases of long ($d=\dup=20 \dpu$) and short ($d=\dup=\dpu$) term protection. Power-law networks with $N=10^5$, $\gamma=3.5$, $\kmin=3$, and $\kmax\sim N^{1/\gamma}$ were used in both cases.}
	\label{fig:limit_case}
\end{figure}

The time in the protected compartment also impacts the epidemic outbreaks. We consider again the limit case $\dpu\ll \dup$ where, once protected, an individual remains in this configuration for a period (vaccination with waning immunity could play this role). Figure~\ref{fig:limit_case} compares the epidemic threshold, prevalence peak, and outbreak duration for long and  short term protection schemes {considering} $R_0^\text{(SIR)}=2.4$. The epidemic prevalence, outbreak duration, and threshold are highly affected by perception rate in the case of long-term protection while the effects are substantially softened for short-term protection. Local awareness is much more effective if protective attitudes are adopted quickly and released slowly, especially in the heterogeneous case where hubs are much less responsive to the local environment.

The targeted immunization strategies in complex networks are, in general, {ruled} by some network centrality~\cite{Satorras2002,Holme2002} or dynamical feature~\cite{Matamalas2018}. The behavioral change as a consequence of local awareness in our model combines dynamical and network features, such as does the nonmassive immunization strategy in which hubs are protected~\cite{Costa2020}. Therefore, in those networks, the behavior response as a consequence of local awareness should assume another form to promote a more effective nonmassive immunization strategy addressing information to hubs~\cite{Bianconi2021}.

\section{Conclusions}
\label{sec:conclusions}
Our capacity to mitigate an emerging epidemic outbreak causing a severe disease without efficient treatments  and to  reduce its impact in a susceptible population depends strongly on the adoption of nonpharmaceutical interventions and, consequently, on the adhesion of the population to adopt the necessary attitudes. However, this involves an adaptive behavior of the population that can respond or not to the call for preventive, and many times unpopular, actions. So, the perception of the epidemiological scenario plays a central role in this decision making. In the present work, we investigate a SIR epidemic model including two susceptible compartments where the respective populations adopt or not preventive behaviors to reduce the contagion risk. Switching between protected and unprotected compartments is driven by the perception of epidemiological situations, quantified by either the local or global epidemic prevalence. We present both a HMF theory and extensive stochastic simulations to tackle the role of adhesion to preventive behavior in the epidemic outbreak and, particularly, on the amount of people which must adopt prevention to efficiently mitigate the epidemic outbreak. 

Our results indicate that local awareness of the epidemic prevalence can effectively reduce the  outbreak size (flattening of the epidemic curve) and increase the epidemic threshold, being more effective for  higher rate of adhesion, when individuals switch  between protected and unprotected compartments. The global epidemic perception, however, is much less effective and not able to alter the epidemic thresholds. These observations are in agreement with previous reports~\cite{HATZOPOULOS2011,Wu2012}. However, we observe that the higher the adhesion rate is, less individuals adopt protection being nevertheless still capable of reducing the outbreak size and increasing the epidemic threshold. The last result  is a counter-intuitive result from the mean-field point of view, but can be understood in terms of the quick response to the local epidemic incidence that blocks local outbreaks. We also observe that heterogeneity of the network of contacts makes the local perception rule less effective when compared to a homogeneous network of similar average degree. This effect occurs because hubs are the most efficient spreaders in the SIR dynamics but concomitantly are less sensitive to epidemic prevalence.

The aforementioned results may impact the way to take into account the interplay between information, adaptive behavior, and the social impact on the epidemic contagion in theoretical and applied epidemiology. We expect that our results can help the search for containment measures which are sustainable from a social perspective and effective in the epidemic mitigation. Since the HMF theory was not able to reckon all central features revealed by the stochastic simulations, a theoretical improvement is necessary. We have investigated the recurrent  message passing approach~\cite{Karrer2010,Bianconi2021}, in which the actual structure of the network is considered. However, {in this case,} the results do not advance substantially with respect to HMF theory and the theoretical quest remains open.

\begin{acknowledgments}
DHS thanks the support given by \textit{Coordenação de Aperfeiçoamento de Pessoal de Nível Superior} (CAPES)- Brazil (Fellowship 465618/2014-6)   and \textit{Fundação de Amparo à Pesquisa do Estado de São Paulo} (FAPESP)-Brazil (Grants no. 2021/00369-0 and 2013/07375-0).
CA thanks the support given by \textit{Conselho Nacional de Desenvolvimento Científico e Tecnológico} (CNPq)-Brazil	(Grant 311435/2020-3) and \textit{Fundação de Amparo à Pesquisa do Estado de Rio de Janeiro} (FAPERJ)-Brazil (Grant   CNE E-26/201.109/2021).
SCF thanks the support by the Brazilian agencies 
CNPq(Grants no. 430768/2018-4 and 311183/2019-0) and \textit{Fundação de Amparo à Pesquisa do Estado de Minas Gerais} (FAPEMIG)-Brazil (Grant no. APQ-02393-18).
This study was also financed in part by CAPES - Finance Code 001.
\end{acknowledgments}

\appendix

\section{Stochastic simulations}
\label{app:algo}
 
The model is implemented considering an optimized Gillespie algorithm adapting the one described in Ref~\cite{Cota2017}. The process of transition from $\Su$ to $\Sp$ compartments with local awareness is implemented as follows. We compute the total transition from $\Su$ to $\Sp$ which is given by  
\begin{eqnarray}
W_{\text{u}\rightarrow\text{p}}  = \dpu\sum_{i\in\Su}\sum_{j}\frac{A_{ij}}{k_{i}}\sigma_{j},
\end{eqnarray}
where $\sigma_{i}=1$ if the neighbor is infected, $\sigma_{i}=0$ otherwise, and the sum runs over the nodes in the compartment $\Su$. Similarly, the total rate associated to the reverse transition $\Sp$ to $\Su$ is given by
\begin{eqnarray}
W_{\text{p}\rightarrow\text{u}}  = \dup\sum_{i\in\Sp}\sum_{j}\frac{A_{ij}}{k_{i}}(1-\sigma_{j}).
\end{eqnarray}
Finally, with the total number of infected nodes given by
\begin{equation}
N_{\text{inf}} = \sum_{j}\sigma_{j}
\end{equation}
and total number of edges emanating from them 
\begin{equation}
N_\text{e} = \sum_{j}k_j\sigma_j\,,
\end{equation}
the stochastic simulations can be implemented as follows. With probability
\begin{equation}
P_\text{heal}=\frac{\mu N_\text{inf}}{\mu N_\text{inf}+\lambda N_\text{e}+W_{\text{u}\rightarrow\text{p}}+W_{\text{p}\rightarrow\text{u}}} 
\label{eq:OGA1}	
\end{equation}
an randomly selected infected node heals and moves to the recovered state. With probability 
\begin{equation}
P_\text{inf}=\frac{\lambda N_\text{e}}{\mu N_\text{inf}+\lambda N_\text{e}+W_{\text{u}\rightarrow\text{p}}+W_{\text{p}\rightarrow\text{u}}}  \,,
\label{eq:OGA2}			
\end{equation}
an infected node $i$ is selected with probability proportional to its degree and one of its neighbors $j$ chosen with equal chance. If $j$ belongs to $\Sp$ or $\Su$, it is infected with  probabilities $\alpha$ and $1$, respectively. No change of state happens otherwise. Finally, with probability  
\begin{equation}
P_{\text{u}\rightarrow\text{p}}=\frac{W_{\text{u}\rightarrow\text{p}}}{\mu N_\text{inf}+\lambda N_\text{e}+ W_{\text{u}\rightarrow\text{p}} +W_{\text{p}\rightarrow\text{u}}} \,,
\end{equation} 
a node $i$ in the $\Su$ compartment  is selected with probability proportional to the fraction of infected neighbors $\omega_i=\sum_{j}A_{ij}\sigma_j/k_i$ and the node is moved to the protected state $\Sp$. Finally, with the complementary probability 
\begin{equation}
P_{\text{p}\rightarrow\text{u}}=1-P_\text{heal}-P_\text{inf}-P_{\text{u}\rightarrow\text{p}},
\end{equation}
a node $i\in\Sp$ is selected with probability proportional to $1-\omega_i$ and moved to the $\Su$ state. Time is incremented by  
\begin{equation}
\delta t  = \frac{-\ln u}{\mu N_\text{inf}+\lambda N_\text{e}+ W_{\text{u}\rightarrow\text{p}} +W_{\text{p}\rightarrow\text{u}}}\,,
\end{equation}
where $u$ is a pseudo random number uniformly distributed in the interval $(0,1)$. 

The algorithm above can be adapted for the global awareness rule using $W_{\text{u}\rightarrow\text{p}}=\dpu N_\text{u}$ and $W_{\text{p}\rightarrow\text{u}}=\dup N_\text{p}$, where $N_\text{p}$ and $N_\text{u}$ are the numbers of protected and unprotected individuals, and selecting the one who is going to change the behavior with equal chance within the corresponding compartment $\Sp$ or $\Su$.

%

\end{document}